# EMICSS: Added-value annotations for EMDB entries

Amudha K. Duraisamy, Neli Fonseca, Gerard J. Kleywegt, Ardan Patwardhan, Kyle L. Morris*

Duraisamy, A. K. - Electron Microscopy Data Bank, European Molecular Biology Laboratory, European Bioinformatics Institute (EMBL-EBI), Wellcome Genome Campus, Hinxton, Cambridgeshire, CB10 1SD, United Kingdom.
Fonseca, N - Electron Microscopy Data Bank, European Molecular Biology Laboratory, European Bioinformatics Institute (EMBL-EBI), Wellcome Genome Campus, Hinxton, Cambridgeshire, CB10 1SD, United Kingdom.
Kleywegt, G. J. - Electron Microscopy Data Bank, European Molecular Biology Laboratory, European Bioinformatics Institute (EMBL-EBI), Wellcome Genome Campus, Hinxton, Cambridgeshire, CB10 1SD, United Kingdom.
Patwardhan, A - Electron Microscopy Data Bank, European Molecular Biology Laboratory, European Bioinformatics Institute (EMBL-EBI), Wellcome Genome Campus, Hinxton, Cambridgeshire, CB10 1SD, United Kingdom.
Morris, K. L. - Electron Microscopy Data Bank, European Molecular Biology Laboratory, European Bioinformatics Institute (EMBL-EBI), Wellcome Genome Campus, Hinxton, Cambridgeshire, CB10 1SD, United Kingdom.

*Correspondence to Kyle L. Morris (kyle@ebi.ac.uk)

# Abstract

**Motivation:**
The Electron Microscopy Data Bank (EMDB) is a key repository for 3D electron microscopy (3DEM) data but lacks comprehensive annotations and connections to most of the related biological, functional, and structural data. This limitation arises from the optional nature of such information to reduce depositor burden and the complexity of maintaining up-to-date external references, often requiring depositor consent. To address these challenges, we developed EMICSS (**EM**DB **I**ntegration with **C**omplexes, **S**tructures, and **S**equences), an independent system that automatically updates cross-references with over 20 external resources, including UniProt, AlphaFold DB, PubMed, Complex Portal and Gene Ontology.

**Results:**
EMICSS (https://www.ebi.ac.uk/emdb/emicss) annotations are accessible in multiple formats for every EMDB entry and its linked resources, and programmatically via the EMDB Application Programming Interface (API). EMICSS plays a crucial role supporting the EMDB website, with annotations being used on entry pages, statistics, and in the search system.

**Availability and implementation:**
EMICSS is implemented in Python and it is an open-source, distributed under the EMBL-EBI license, with core code available on GitHub (https://github.com/emdb-empiar/added_annotations).

**Contact:** kyle@ebi.ac.uk

# Introduction

The Electron Microscopy Data Bank (EMDB; emdatabank.org | emdb-empiar.org) stands as the single global repository for 3D volumes and associated data derived from cryogenic-sample electron microscopy (cryo-EM) experiments (wwPDB Consortium, 2024). Established at EMBL-EBI in 2002 (Tagari *et al.*, 2002), EMDB has evolved into a key archive for structural biology using cryo-EM. In 2021, the archive joined the worldwide Protein Data Bank (wwPDB; wwpdb.org) (Berman *et al.*, 2003, 2007; wwPDB consortium, 2019) and it is now developed and managed jointly by the wwPDB partners.

EMDB archives data from various cryo-EM techniques, including single-particle analysis, electron tomography, subtomogram averaging, helical reconstruction, and electron crystallography. The field went through a "resolution revolution" (Kühlbrandt, 2014; Bai *et al.*, 2015) around 2014, marked by technological advancements including but not limited to the introduction of fast direct electron detectors. This revolution has propelled exponential growth in cryo-EM, mirrored by a substantial growth of EMDB and associated models in the PDB as illustrated in **Figure 1**.

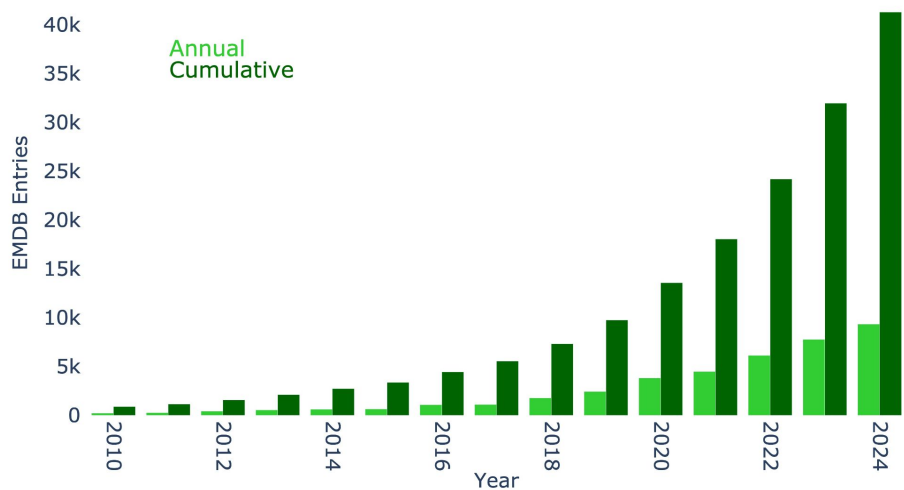

**Figure 1**. Annual (light green) and cumulative (dark green) EMDB entries released until December 2024.

The wwPDB partners maintain and develop the OneDep system (https://deposit-2.wwpdb.org/deposition) that facilitates the deposition and validation of atomic coordinates (PDB), cryo-EM maps (EMDB), or NMR experimental data (BMRB). Structure depositors can use the web-based deposition interface of OneDep to upload data and provide metadata related to a structure experiment. Data items relevant to EMDB include the primary map representing the main outcome of the cryo-EM experiment which is typically used to generate images shown in the associated publication, and the half maps used for validation. Metadata items relevant to EMDB include links to PDB models that may have been fit to the cryo-EM map, citation information, author details, sample information, imaging and structure determination, validation criteria, and more. Most of this information is optional to minimise the burden of data entry, in particular links to external databases. Further, even if the information is provided during deposition, maintaining up-to-date external reference information requires substantial effort, and necessitates frequent releases of updated versions for the EMDB entries involved, potentially also requiring depositor approval for the updates. Our approach has therefore been to develop an independent resource - EMICSS (**EM**DB **I**ntegration with **C**omplexes, **S**tructures and **S**equences) which attempts to automatically maintain up-to-date cross references between EMDB and other pertinent resources without directly modifying EMDB entries, which are strictly maintained to reflect what was provided by a depositor and curated by the wwPDB.

Enriching entries with robustly obtained and rich external database links improves archived data utility reuse beyond their primary study. Specifically, the Findability, Accessibility, Interoperability, and Reusability (FAIR) are crucial for archived data and metadata value to be realised (Wilkinson *et al.*, 2016). Added-value annotations play a crucial role in data Findability by increasing the relevant accession identifiers describing archived entries. These annotations enhance the discoverability of specific information within EMDB by indexing it in the search system, facilitating access to relevant data for scientific researchers, database users, scientific and machine learning developers. EMICSS annotations are made accessible via downloadable files in commonly used formats, as well as through the EMDB search system, API endpoint and entry pages, and are freely available to all users. Standard terms, such as Gene Ontology (GO) for describing molecular functions, biological processes, and cellular components, or qualified references, such as accession identifiers used by other databases/resources are employed with the aim of maintaining interoperability and potential integration of EMDB entries with external resources. Finally, EMICSS metadata is enriched with detailed provenance information, information about the origin of the data, facilitating the appropriate utilisation of annotated EMDB entries in other archives, meta studies, and analyses by researchers and other databases, fostering a culture of data reuse. Essentially, EMICSS annotations build and reinforce the FAIR usage of EMDB.

In this paper we describe the EMICSS pipeline which generates a comprehensive set of added-value annotations for EMDB entries. The generated data is gathered by mining data in external bioinformatics resources to enhance the cross-referencing of EMDB entries. The PDBe/UniProt-SIFTS (**S**tructure **I**ntegration with **F**unction, **T**axonomy and **S**equences resource) (Velankar *et al.*, 2013; Dana *et al.*, 2019) operates analogously to maintain up-to-date annotations for PDB model entries, principally targeting protein chains within PDB entries to facilitate cross-referencing at the residue level. EMICSS generates cross references at various levels (entry/sample/sequence) whether the entry is map-only or has associated model coordinates in the PDB. These entry-level cross references differ from other cross referencing tools such as SIFTS from PDBe, which cross references at model coordinate residue level. This is particularly important as not all EMDB entries have fitted models, and even when models are available, they may not represent the entire map. Therefore, unlike SIFTS, EMICSS aims to provide cross-linking at non-residue levels. The EMICSS workflow, processes, integrates, filters, and transforms information to generate and provide linked data outputs intended to cater for different classes of users; website users, bioinformaticians, data re-users, and other biological resources. The enrichment of EMDB data facilitated by EMICSS provides additional data describing an entry, aimed at improving the search process and chances of data discovery. EMICSS data allows external biological resources to link or reference EMDB entries from their own resources and data. EMICSS operates as a fully automated pipeline regularly updating the added-value annotations alongside the weekly release and update of EMDB entries.

## System and methods

EMICSS automatically generates annotations for new entries in the weekly EMDB release and updates annotations for the whole EMDB archive. This ensures that annotations for every EMDB entry remain up-to-date, and new annotations are added for the latest releases. **Table 1** lists the primary resources queried by EMICSS to gather annotations across entry level, sample level, and sequence level.

The EMICSS annotation generation workflow consists of four main steps, as outlined in **Figure 2**. The "Data preparation" step involves data harvesting from other external resources. For example, UniProt IDs associated with a PDB entry file are downloaded and used to match entries in Complex Portal, ensuring that the data inputs to EMICSS include the most recent updates. The second step collects added-value annotations for EMDB entries where annotations are collected on an entry-by-entry basis. The annotation collection process starts by extracting cross-references stored in the entry's header file which may contain some cross-references that were provided at the time of deposition. In cases where conflicting annotations arise between depositor-provided data and external sources (e.g., a DOI provided by the depositor does not match the one from PubMed), the depositor-provided data is trusted and used. Annotations from external sources are included only when the corresponding information is missing in the depositor's metadata. Provenance in EMICSS is an attempt to highlight whether such conflicts exist, as this could be useful for users to understand the source of the annotated data. The collected annotations are then written into separate EMICSS TSV resource files; these files contain all annotations for all entries within the archive. The third step generates EMICSS XML files containing details of all cross-references on an entry-by-entry basis. Both the TSV and XML files are EMICSS files are public available at EBI FTP. In the final step, the EMICSS data is loaded into a graph database, providing input to the Annotations API which is subsequently used by the entry pages to show the annotations. Additionally, the data is indexed for the EMDB search engine, enhancing the functionality of the search API, search pages, and EMDB statistics.

**Table 1.** Resources utilised for annotations at entry, sample and sequence levels, respectively.

| Entry-level annotations | PDB (wwPDB consortium, 2019), EMPIAR (Iudin *et al.*, 2016, 2023),, Europe PubMed Central (Roberts, 2001), DOI, ISSN (www.issn.org) and ORCiD (Butler, 2012) |
|---|---|
| Sample-level annotations | Complex Portal (Meldal *et al.*, 2019), UniProtKB (UniProt Consortium, 2021), ChEMBL (Mendez *et al.*, 2019) , ChEBI (Hastings *et al.*, 2016) , DrugBank (Wishart *et al.*, 2018), PDBe-KB (PDBe-KB consortium, 2020), AlphaFold DB (Jumper *et al.*, 2021; Varadi *et al.*, 2022) and Rfam (Kalvari *et al.*, 2021) |
| Sequence-level annotations | Gene Ontology (Ashburner *et al.*, 2000; The Gene Ontology Consortium, 2021), Pfam (Mistry *et al.*, 2021), InterPro (Blum *et al.*, 2021), CATH (Sillitoe *et al.*, 2021), SCOP (Andreeva *et al.*, 2020), SCOP2 (Lo Conte *et al.*, 2000; Andreeva *et al.*, 2014) and SCOP2B (Andreeva *et al.*, 2014) |

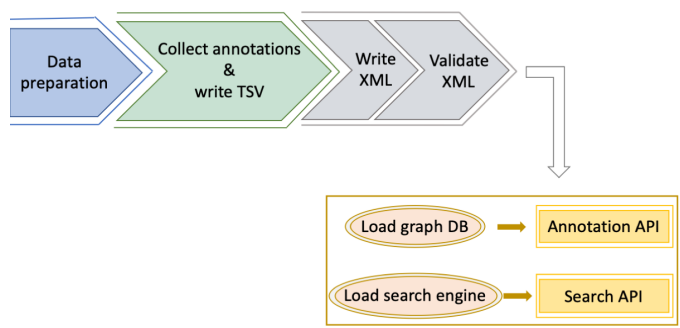

**Figure 2.** Summary of the key steps in the EMICSS annotation generation protocol.

# Algorithm

The process of acquiring added-value annotations that the authors have not supplied at the time of deposition utilises mapping entries to various resources as described, various methods are employed to collect information from diverse resources (**Figure 3**), outlined as follows.

### 3.1. Structures and protein annotations

#### 3.1.1. EMPIAR
The associated EMDB accession identifiers are obtained from the EMPIAR metadata files. EMPIAR mapping is done for the whole EMDB archive at once rather than on an entry-by-entry basis, as entry-by-entry mapping would require reading all metadata files repeatedly for each mapping.

#### 3.1.2. UniProt
The data retrieved from Uniprot REST API endpoint serves as a resource for harvesting data that contains mappings between UniProt IDs, their corresponding PDB IDs, and the protein names for the whole archive. For EMDB entries with associated PDB models, the corresponding UniProt IDs are assigned by conducting a broad search that finds matches even when the search query doesn't perfectly match corresponding data. The UniProt protein name is compared with the author-provided protein name. Broad searching allows EMICSS to identify matches even in instances of slight discrepancies or variations in spelling. However, a Uniprot mapping may fail if the protein name provided by the depositor is not good enough to match with the list of UniProtKB protein names. For entries that lack a PDB ID, and if the sequence information is provided by the depositor, a protein sequence search using BLASTP is performed to identify the relevant UniProt IDs for the proteins in the entries.

#### 3.1.3. PDB
It is a requirement of most reputable journals that for structural biology publications, if any atomic coordinates have been modelled into a cryo-EM map that these are deposited to the EMDB, and that modelled atomic coordinates should be deposited to PDB. Consequently, the accession identifiers of the modelled coordinates are stored in the EMDB header file and can be used to extract relevant information from the PDB archive, SIFTS, PDBe assemblies and other resources. For each PDB model, the molecular weight of the molecule or complex represented by that PDB entry and the number of copies of the biological assemblies in the sample are extracted from the corresponding PDBe's FTP assembly file. We note that inconsistencies between assemblies in the PDB and those in EMDB, or if the model represents only a part of an EM map, could lead to inaccuracies in calculating the overall molecular weight.

#### 3.1.4. Complex Portal
The previously identified UniProt IDs are used to link complexes in the sample to entries in the Complex Portal using mapping files available from the Complex Portal (https://ftp.ebi.ac.uk/pub/databases/intact/complex/current/complextab/). This annotation process also includes calculating a matching score for each mapping. EMICSS compares both the complete complex and its sub-complexes and a matching score is generated to evaluate the degree of similarity between them. To calculate the score, the UniProt identified complex associated with the EMDB entry (A) is compared to Complex Portal's complexes via their UniProt IDs (B). A comparison score is determined by the Jaccard index, defined as the ratio of the size of the intersection (A∩B) to the size of the union (A∪B). A score of 1.0 indicates an exact match, which equates to a definitive mapping. Only complexes with a score above 0.5 are considered, and all these complexes are listed. Errors may occur in mapping the Complex Portal if a group of proteins is expected to form a complex but is not observed as such in the entry. Also, it is a challenge to map the complexes when a set of proteins can be associated with multiple Complex Portal IDs.

### 3.1.5. AlphaFold DB
References to AlphaFold DB (AFDB) are acquired from the AlphaFold DB-UniProtKB mapping available at AFDB FTP area (https://ftp.ebi.ac.uk/pub/databases/alphafold/accession_ids.csv). This may be particularly useful for proteins that have not been modelled (or only partly modelled) by the depositors.

## 3.2. Citation annotations
### 3.2.1. PubMed
This information is solely provided by the authors, and EMICSS does not utilise any mappings to obtain the PubMed IDs.

### 3.2.2. ORCiD, PMC, DOI, ISSN
When there is a provided PubMed ID, the Europe PMC (Ferguson *et al.*, 2021) search API is used to retrieve comprehensive citation information including ORCiD IDs, PubMed Central (PMC) ID, Digital Object Identifier (DOI) and International Standard Serial Number (ISSN).

## 3.3. Ligand - ChEMBL, ChEBI & DrugBank
The information about ligands in the EMDB is extended by utilising the wwPDB Chemical Compound Dictionary (CCD) (Feng *et al.*, 2004). In this process, the wwPDB ligand atom format known as HET is employed to find corresponding ChEMBL, ChEBI, and DrugBank identifiers.

## 3.4. RNA - Rfam
The PDBe mapping of RNA is leveraged to link an EMDB entry to Rfam. By utilising the PDB ID, the corresponding Rfam accession can be identified. When there are multiple RNAs associated with a single PDB ID, the process uses the sample ID - an ID employed to describe the entry's sample in the EMDB metadata files - to accurately map each RNA to the correct Rfam accession.

## 3.5. Sequence annotations- GO, InterPro, Pfam, CATH, SCOP, SCOP2B
With the availability of atomic coordinates and sequence information in the EMDB entry, annotations from Gene Ontology (GO), InterPro, Pfam, CATH, SCOP, SCOP2, and SCOP2B are extracted from the respective PDBe/UniProt SIFTS file. The inclusion of sequence information is crucial to annotate only those terms and domains relevant to the region of the protein represented by the structure. The names and classifications of terms within the GO hierarchy, along with the broader biological categories to which they belong are retrieved using the UniProt API (https://rest.uniprot.org/uniprotkb/{uniprot_id}.xml). Additionally, Smith–Waterman algorithm, a local alignment is conducted between the author-provided and PDB sequences to identify matching residues.

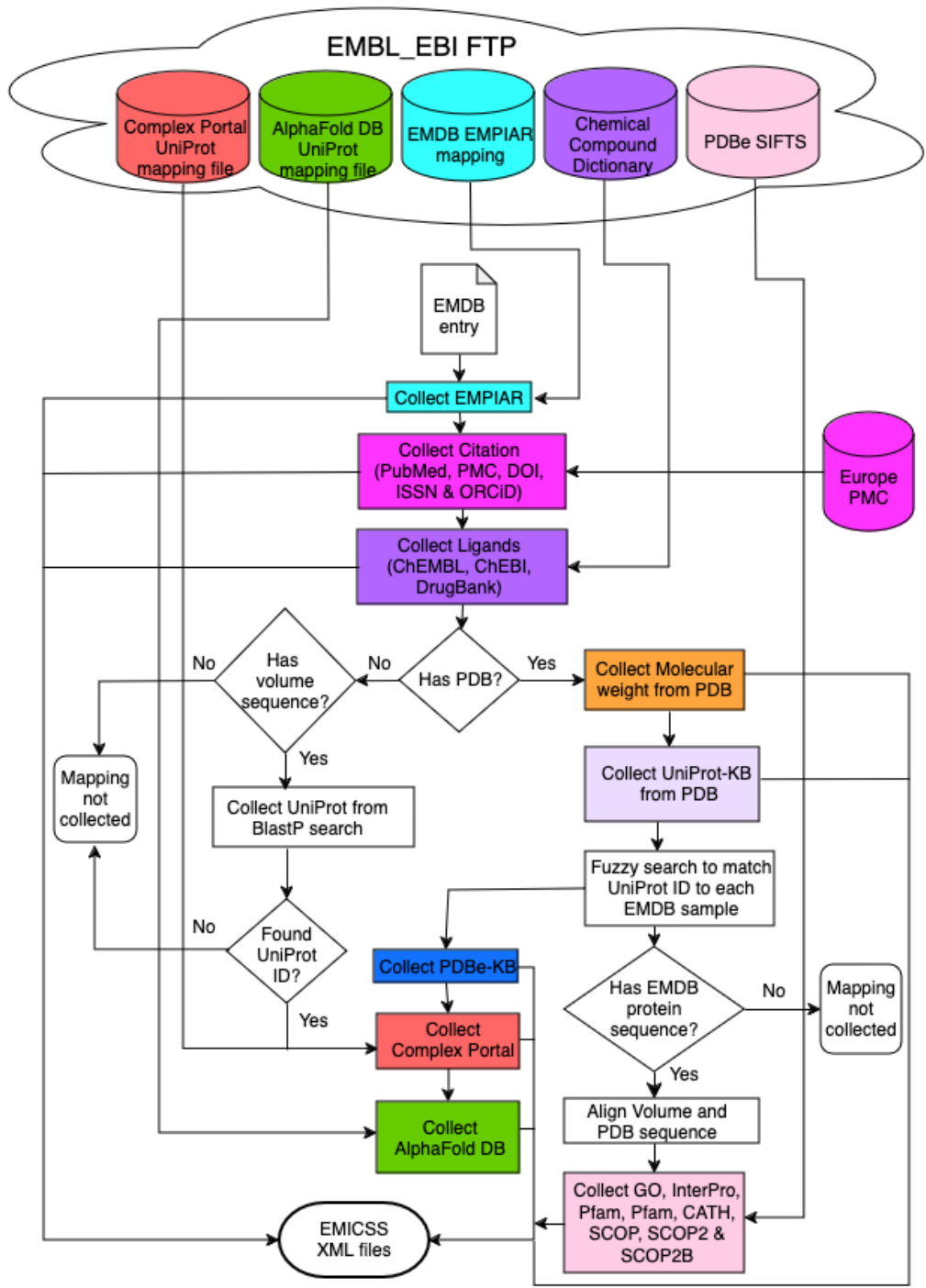

**Figure 3.** "Flowchart depicting the protocol used in the collection step of the EMICSS pipeline (labeled as 'collect annotations' in Figure 2), which is employed to map various resources to EMDB entries. The next step in the pipeline is 'write TSV'.

# Implementation

The workflow of EMICSS produces various components aimed at enhancing utility and accessibility of EMDB entries, these data are disseminated and applied in a number of ways as follows. EMICSS files are made publicly available and serve as comprehensive mapping information for both resources and entries. Additionally, the EMDB search system plays a pivotal role in utilising EMICSS data through its implementation of auto suggestion while typing, facets, filters, and statistical tools. Furthermore, the integration of added-value annotations directly into EMDB entry pages enhances entry contextualisation. Finally, the availability of an EMICSS annotation API endpoint (https://www.ebi.ac.uk/emdb/api/annotations/) offers programmable access to EMICSS data, enabling developers and researchers to incorporate contextual information into their workflows and applications should they choose.

## 4.1. EMICSS files

EMICSS produces annotations in two formats: resource-based files, as individual per-resource TSV files containing all annotations for EMDB entries pertaining to that resource and entry-based, as individual per-entry XML files containing all annotations available for an EMDB entry from all resources. Both the TSV and XML files are available from the EMDB-EMICSS download link (https://ftp.ebi.ac.uk/pub/databases/em_ebi/emdb_related/emicss/). The XML files are also accessible from individual EMDB entry pages, in the Download menu and in the metadata section of the Links tabs.

### 4.1.1. TSV files

EMICSS produces annotations in TSV format on a per-resource basis. We currently distribute 19 TSV files, one for each resource, e.g. UniProt, AlphaFold DB, Complex Portal, etc. Each file is structured with columns relevant to the respective resource. Common elements across all files include the EMDB ID, the mapping resource accession ID, provenance (indicating the source of this mapping), and additional columns specific to the resource type.

### 4.1.2. XML files

EMICSS produces annotations in XML format for each EMDB entry. The XML data model, EMDB_EMICSS.xsd
(https://ftp.ebi.ac.uk/pub/databases/em_ebi/emdb_related/emicss/emicss-schema/current/),
is used to generate EMICSS XML files. These files encompass information on XSD versioning, data-model location, and the EMDB ID. If there is a database release version for a resource used in EMICSS, this information is also included in the XML files. Additionally, the XML files include the date of all annotations added to EMICSS for that specific entry. To account for potential updates to external resources that are cross referenced, EMICSS is able to automatically regenerate all XML files for all entries of the EMDB archive as required.

### 4.1.3. Usage by other resources - UniProt, EMDR, Europe PMC

EMICSS TSV files serve as a valuable resource for cross-referencing EMDB entries with other relevant datasets. For instance, the EMDataResource (EMDR) uses the EMICSS file to cross-reference EMPIAR data associated with EMDB entries. UniProt leverages the specifically formatted EMICSS file to cross-reference EMDB entries in the UniProt-KB resource. Additionally, EMICSS file is used to cross-reference EMDB entries in Europe PMC.

### 4.2.4. Intended users

EMICSS is targeted at a varied user base through various different channels. Driving some information found on EMDB entry pages, statistics, API and the search system EMICCS enables website visitors to derive insights, discover linked resources and characterise entry properties. At the same time, other databases in the same domain can leverage EMICSS data to map EMDB entries, establishing connections between resources. This interconnected approach simplifies information retrieval for users by consolidating relevant data in a single resource.

## 4.2. EMDB search system

EMICSS data are indexed in the EMDB search engine. This integration broadens the scope of query possibilities and thus data findability. Users can now discover macromolecular structures by referencing Gene Ontologies, or explore specific domains like Zinc Finger and PDZ domains extracted from InterPro and Pfam. Moreover, the service facilitates identifying entries associated with particular ligands through standardised nomenclature drawn from ChEMBL, ChEBI, and Drugbank. Notably, EMICSS indexing enables queries using external resource identifiers, thus increasing structural data discovery. For instance, entries associated with a specific biological process such as GTP metabolic processes will have an EMICSS annotation for that process's Gene Ontology identifier. Thus searching by a specific GO identifier (GO:0046039) or by typing the process name will identify entries associated with that biological process. This search feature also offers auto-suggestions while typing making it even easier to select the desired term. Another example includes making queries for entries with associated ChEMBL data. Furthermore, this indexing enhances the statistical insights derived from EMDB search results or through the EMDB Chart Builder tool (https://www.ebi.ac.uk/emdb/statistics/builder), equipping users with comprehensive data analysis capabilities.

## 4.3. Entry page annotations

Annotations are accessible in different tabs of the individual EMDB entry pages, such as the Overview, Sample, and Links tabs. The specific details of annotations in each of these tabs are elaborated below.

### 4.3.1. Overview tab

In the Overview tab, annotations based on the citations are included. If these details were not provided during deposition, EMICSS supplements the information with ORCiD IDs of the authors, PMCID, DOI, ISSN and ASTM (American Society for Testing and Materials) code of the publication, all obtained using the PubMed ID of a publication.

### 4.3.2. Sample tab

EMICSS data increases the amount of information that can be shown in the Sample tab. This includes information such as the overall molecular weight of the sample and cross-referencing to complexes, proteins, and ligands if they are present in the sample. Complex annotations encompass the first few entries from the list of annotated complexes. These details are presented in a table that includes corresponding accession IDs and complex names from the Complex Portal, along with their respective overlap scores.

Proteins are connected to their respective entries in UniProt, AlphaFold DB, GO (inclusive of GO categories), InterPro, Pfam, CATH, SCOP, and SCOP2B, with reference to the structural sequence. Ligands are associated with ChEMBL, ChEBI, and DrugBank accession IDs. RNAs are associated with Rfam accessions. All annotated accession IDs are hyperlinked, facilitating easy navigation to the corresponding entry pages in external resources. **Figure 4** shows the sample tab for entry EMD-0403.

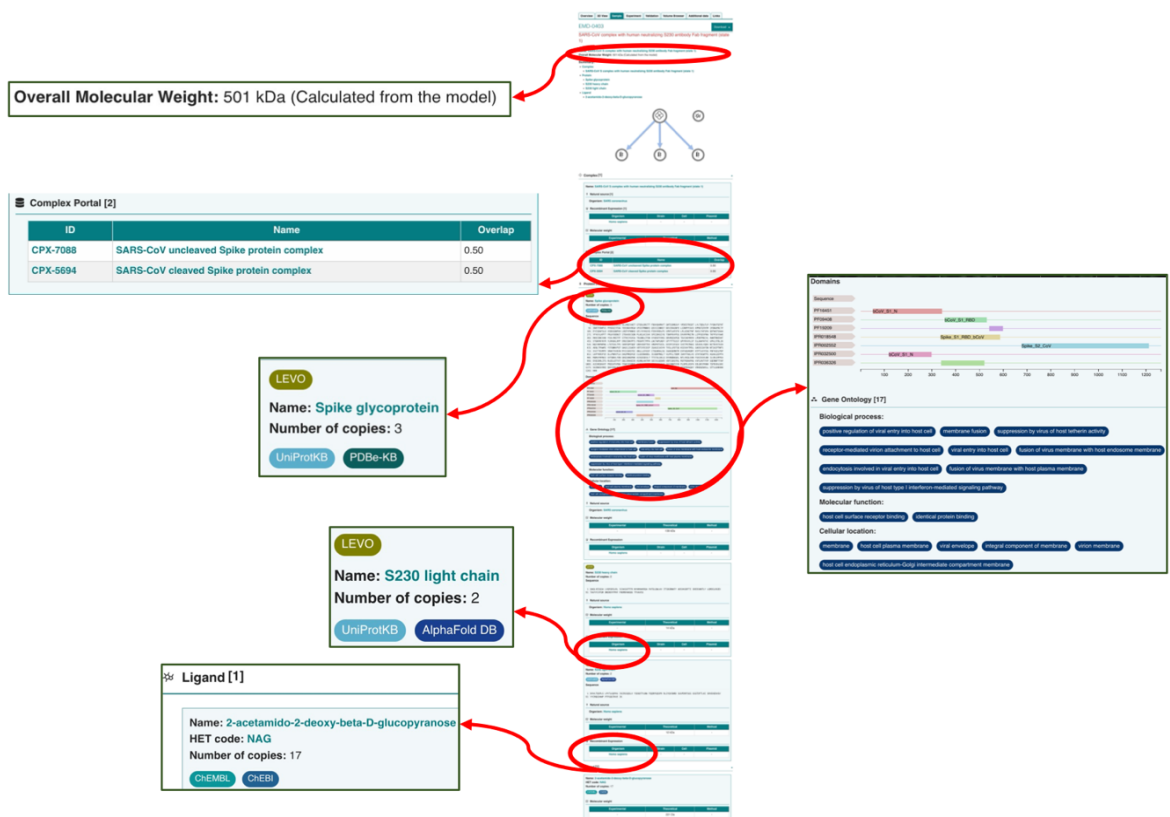

**Figure 4.** Illustration of EMICSS data within the "Sample" tab of the EMD-0403 entry. The highlighted red circles represent enrichments performed by EMICSS.

### 4.3.3. Links tab

In the Links tab, the Similar Entries table is exclusively sourced from EMICSS. These entries are identified through neighbourhood overlapping, calculated using the Jaccard coefficient in the graph database. If similar entries exist for a particular entry, they are listed. An example entry where the Similar Entries table can be found is EMD-8117.

### 4.4. EMDB Annotation API

The Annotation API consists of three distinct endpoints, each serving a different purpose in enhancing the functionality and information retrieval in EMDB.

### 4.4.1. Entry annotation endpoint

This endpoint provides detailed annotations for an EMDB entry, which is used to populate the Sample tab on EMDB entry pages. Additionally, end-users can leverage this endpoint to retrieve all annotations linked to an entry in JSON format. Endpoint: https://www.ebi.ac.uk/emdb/api/annotations/

### 4.4.2. Similar entries endpoint

This endpoint returns entries similar to a given EMDB entry by finding the neighborhood overlapping nodes in the graph database. Similar entries are listed in Links tab of an entry page as mentioned in section 4.3.3
Endpoint: https://www.ebi.ac.uk/emdb/api/emicss/similarities/

### 4.4.3. EMICSS release statistics endpoint

The third endpoint is designed for obtaining EMICSS statistics for the latest release. It serves as a resource for users seeking an overview of the EMICSS release, which includes the distribution of entries across different resources, contributing to a better understanding of the EMICSS dataset, as depicted in **Figure 5**.
Example endpoint: https://www.ebi.ac.uk/emdb/api/emicss/releases

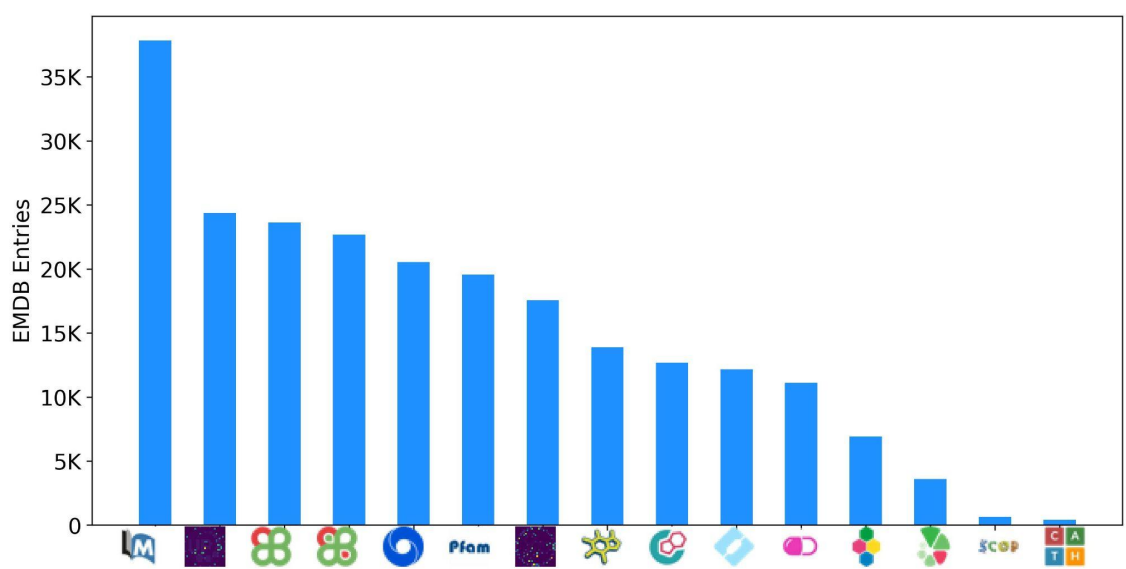

**Figure 5.** Displays the release statistics of EMICSS acquired through the annotation API.

### 4.5. Potential future uses

#### 4.5.1. Opportunities from exposing database links

AlphaFold models are identified by EMICSS, including for map-only entries. The number of map-only entries that lack an associated model from the primary deposition is small, as shown in figure 6. However, the identification of an AlphaFold model for map-only entries opens up the opportunity for proposing models which may appropriately represent unmodelled cryo-EM density.

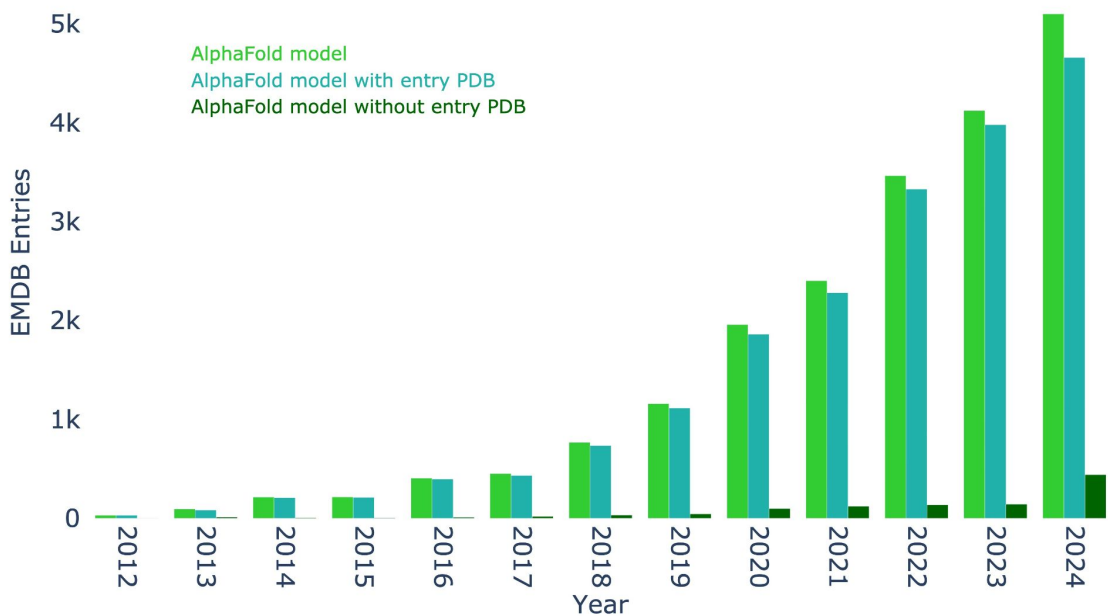

**Figure 6.** The total number of EMDB entries with an identified AlphaFold model. Entries with an identified AlphaFold model may also have associated model coordinates deposited to the PDB or may not have associated model coordinates.

## Discussion

The EMICSS annotation generation pipeline operates in synchrony with the weekly EMDB release. Every step is automated, from annotating, mapping, and backing up collected annotations, acquiring necessary EMICSS XML and TSV files, placing them in the FTP, loading annotated data into relational and graph databases (used for search and the annotation REST API). External resources can utilise publicly available EMICSS data to map EMDB entries to their own database entries. The entry-wise mapping file enables access to comprehensive annotation information for a single entry, while entries sharing a common annotation term can be searched and downloaded as a unified file from the EMDB webpage. EMICSS data is deeply embedded and accessible on each entry page. Benefited by the enrichment of metadata of EMDB entries, EMICSS enables more sophisticated search queries to be made of the archive. More fundamentally, a richer and user-friendly experience is delivered on the EMDB website with the addition of EMICSS data. Where the annotation process is fully automated, and some associations are the result of broad searches we expect some false positive identification but overall EMICSS is expected to increase the chances of data discovery. Detailed documentation and information can be accessed on the dedicated EMICSS landing page (https://www.ebi.ac.uk/emdb/emicss). This page provides access to download links and statistics, offering a comprehensive view of the EMICSS coverage for each resource in the latest release and tracking the development of this coverage over time. This annotation enrichment workflow is centrally important to reinforce the FAIRness of data in the EMDB archive and builds further infrastructure to extend the biological knowledge derivable from a structure beyond that of the primary studies conclusions. The EMDB aims to extend the concept of post-publication added-value annotation to further enrich the information in the archive.

## Funding

This work was supported by the European Molecular Biology Laboratory–European Bioinformatics Institute; and the Wellcome Trust [grant number 212977/Z/18/Z].

## Acknowledgements

We are grateful to our EMDB and EMPIAR colleagues for the help and support provided throughout this project, and in particular: A. Iudin for bug fixes and comments, J.Turner, P. Korir, S. Abbott, R. Pye, and Z. Wang for help and valuable feedback on the project and manuscript. We are also grateful to several members of the PDBe team with the PDBe-SIFTS resource and files, and in particular to S. Valenkar, P. Choudhary, S. Anyango and J. Berrisford. We would also like to thank T. Goddard (UCSF), E. Peisach (RCSB-PDB), J.M. Carazo (3DBionotes) and the PDBj team (Osaka) for feedback on earlier incarnations of the EMICSS files.

## Code availability

Our code is available at https://github.com/emdb-empiar/added_annotations

## Declaration

We acknowledge the use of ChatGPT (OpenAI) as a tool to assist with spell checking and grammatical corrections during the preparation of this manuscript. All intellectual content, methods, codes, interpretations, and conclusions are the sole responsibility of the authors.